\documentclass[12pt,preprint]{elsarticle}
\usepackage[latin9]{inputenc}
\usepackage{color}
\usepackage{amsmath}
\usepackage{amssymb}
\usepackage{graphicx}



\begin{document}
\begin{frontmatter}

\title{Emergence of Brain Rhythms: Model Interpretation of EEG Data\tnoteref{t1}}

\tnotetext[t1]{All authors equally contributed to write the manuscript.}

\author[rvt]{Javier A. Galadí\fnref{pa,ss}}

\author[rvt]{Joaquín J. Torres\corref{cor1}\fnref{dr,ss}}

\author[rvt]{J. Marro\fnref{dr}}

\address[rvt]{Institute ``Carlos I'' for Theoretical and Computational Physics,
University of Granada, Granada, Spain. }

\fntext[pa]{Current Address: Department of differential equations and numerical
analysis, University of Seville, Spain.}

\fntext[dr]{These authors proposed and designed the research.}

\fntext[ss]{These authors wrote the codes and performed the simulations.}

\cortext[cor1]{Corresponding author: jtorres@onsager.ugr.es}
\begin{abstract}
{\em Electroencephalography} (EEG) monitors ---by either intrusive
or noninvasive electrodes--- time and frequency variations and spectral
content of voltage fluctuations or waves, known as \textit{brain rhythms},
which in some way uncover activity during both rest periods and specific
events in which the subject is under stimulus. This is a useful tool
to explore brain behavior, as it complements imaging techniques that
have a poorer temporal resolution. We here approach the understanding
of EEG data from first principles by studying a networked model of
excitatory and inhibitory neurons which generates a variety of comparable
waves. In fact, we thus reproduce $\alpha$, $\beta,$ $\gamma$ and
other rhythms as observed by EEG, and identify the details of the
respectively involved complex phenomena, including a precise relationship
between an input and the collective response to it. It ensues the
potentiality of our model to better understand actual mind mechanisms
and its possible disorders, and we also describe kind of \textit{stochastic
resonance }phenomena which locate main qualitative changes of mental
behavior in (e.g.) humans. We also discuss the plausible use of these
findings to design deep learning algorithms to detect the occurence
of phase transitions in the brain and to analyse its consequences.
\end{abstract}


\end{frontmatter}

\section*{Introduction}

There has been a growing interest in investigating the occurrence
of phenomena associated with thermodynamic-like phase transitions
and criticality during the functioning of neural media by means of
novel experimental techniques, analysis of available connectome data,
and biological-inspired theoretical approaches; see, e.g., \citep{chialvo_eguiluz05,Hesse2014,Torres2015,MC2017,munoz2018},
and references therein. In particular, a sort of brain critical behavior
---mimicking essential features of phase transition phenomena such
as condensation and ferromagnetism--- is now believed to be at the
origin of the observed good processing throughout the brain of signals
coming from different areas and the senses \citep{cocchi2017,gollo2017,munoz2018}.
That is, there has recently emerged definite evidence that weak signals
are optimally transferred and even enhanced in a noisy environment
when the system is in a well-defined region with great susceptibility
which happens to separate neuron dynamic ``phases'', i.e., areas
in parameter space in which the brain shows qualitatively different
kinds of behavior \citep{Torres2015,MC2017}. Ref. \citep{Torres2015}
also presents a feasible procedure to experimentally detect phase
transitions and their details during the performance of actual brains.
Following this promising path, in the present paper we investigate
the possibility of visualizing phase transitions during brain operation
by using easily-extracted brain-activity data obtained from EEG (by
the same token, magnetoencephalograph) recordings. It ensues what
we hope is a convenient tool to monitor \textit{in vivo} changes between
different dynamic behaviors of the cerebral activity. It may also
follow how to design specific stimuli to control these dynamic phases
and eventually modify some of their properties, e.g., in cases of
dysfunction.

More specifically, we here present and discuss an EEG neural-activity
model which generalizes and formalizes a previous one \citep{LopesdaSilva1974}.
We link this to a familiar mathematical framework, improve the temporal
precision of the original setting, include an appropriate tuning of
the noise, and consider the possibility of an input signal that makes
the model useful to reveal and analyze new intriguing phenomena. The
new setting allows us to deep on how oscillation patterns, e.g., as
observed by electroencephalography, emerge reflecting different dynamic
activity, and we thus infer the precise role of the intrinsic noise
in causing familiar rhythms in the human brain. It ensues that not
only $\alpha$ rhythms but also $\beta$, $\gamma$ and ultrafast
oscillations are all just a form (at different levels) of the same
``noise''\ as it is \textit{filtered} (in a way that our model
clarifies) by the neural network itself. That is, one may conclude
that the cause for brain waves is universal within this context, which
allows us to consider a unique mechanism for any of the mentioned
voltage fluctuations with only a relevant parameter. This, we show,
is the intensity of the sum of all the inputs, either noisy or constant,
reaching the network from the outside, and we succeed in parametrizing
it. Consequently, we are able to rigorously relate the occurrence
of phase transitions ---actually, having a non-equilibrium nature
\citep{MD2005,MC2017}--- in the brain with different possible dynamic
behaviors which are revealed by the easily-observed EEG rhythms mentioned.
It is with this aim that we here use a network, which involves both
excitatory and inhibitory units, where a random input is sufficient
to generate different brainwaves, some of them respectively corresponding
to $\alpha$, $\beta$, $\gamma$ and ultrafast oscillations. We also
precisely relate the intensity of the input and the frequency of the
resulting dynamic response and ---following a method first reported
in \citep{Torres2015}---we show how to use an external signal in
a simple experiment to identify the undergoing phase changes and other
details during brain operation using the mechanism of stochastic resonance
(SR) \citep{SRreview98}.

We believe there are two extra, side results from the present model.
One is that it may be useful to design appropriate deep or machine
learning neural networks to learn about possible phase transitions
in the brain and their features -- including critical exponents and
universality classes -- from raw data in actual EEG recordings \citep{zhang2018}.
Also our results here can be used to built similar algorithmic tools
based instead on the SR phenomenon \citep{sra2016} to optimally learn
feature representation in the presence of noise from EEG time series
\citep{deeplearningreview2019}.

%\section*{Materials and methods}

\section*{Model and method}

\begin{figure}[tbh]
\centering{}\includegraphics[width=12cm]{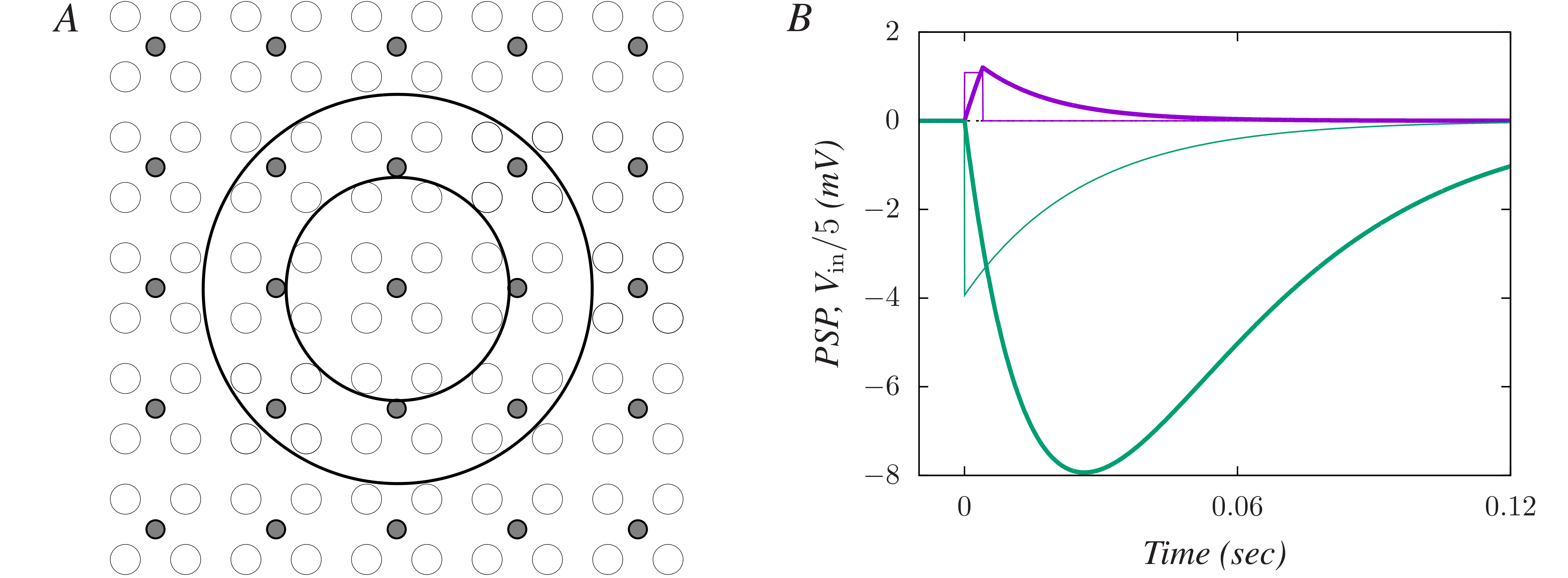}\caption{Model features. \underline{Left panel}: a portion of the actual network
topology (we here in practice considered $N=180$ nodes), where filled
circles stand for inhibitory (\textit{I}) neurons and open circles
represent excitatory (\textit{E}) neurons. In order to mimic biological
conditions (and following \citep{LopesdaSilva1974}), the largest
of the two concentric circles drawn includes 32 \textit{E}s which
influence the \textit{I} at its center, and the smallest concentric
circle includes 12 \textit{E}s under the influence of that \textit{I}.
\underline{Right panel}: an excitatory postsynaptic potential (EPSP;
topmost, purple curve) and an inhibitory one (IPSP; lowermost, green
curve) as modeled using the time-dependent voltage functions $V^{E}(t)$
and $V^{I}(t)$ (see main text) for parameter values in \citep{LopesdaSilva1974},
namely, $t_{\text{max}}=4\,ms,$ $\epsilon=0.3425\,V/s,$ $\eta=-820\,V/s,$
$\tau_{1}=16\,ms$ and $\tau_{2}=26.3\,ms.$ For illustrative purposses,
we also show here (with thinner lines) the two functions in Eq.(\ref{eq:2bis}).}
\label{fig1} 
\end{figure}

Consider, for simplicity and ease of representation, a regular two-dimensional
network on a torus ---in which periodic boundary conditions avoid
surface effects and simulate a larger system--- with $N$ nodes each
holding either an \textit{excitatory }(\textit{E}) or an \textit{inhibitory
}(\textit{I}) neuron as depicted in Fig.\ref{fig1}. They interact
with each other such that any \textit{E} excites one or more \textit{I}s
as long as the membrane voltage of the former exceeds a given threshold
potential, and when any \textit{I} exceeds its own threshold it will
inhibit a group of \textit{E}s (negative feedback). No delays are
considered, which might exclude very extensive networks, and we also
neglect both positive feedback of \textit{E}s (any \textit{E} stimulating
another \textit{E}) and negative feedback of \textit{I}s (any \textit{I}
inhibiting other \textit{I}). Furthermore, according to histological
data ---showing that, in portions of the cortex, there are about
four times more excitatory than inhibitory neurons \citep{Tombol1967,Braitenberg1998,Markram2004,Okun2009}---
the \textit{E/I} ratio is assumed here to be 4, so that any \textit{I}
neuron receives effective excitatory inputs from 32 surrounding \textit{E}s
and any \textit{I} neuron projects upon 12 surrounding \textit{E}s,
as illustrated in Fig.\ref{fig1}A. 

\subsection*{Dynamics}

Each neuron is fully characterized by a potential or ``voltage membrane''\ $V$
which evolves in time ---below a given threshold for firing, $V_{\text{th}}$---
according to a type of integrate-and-fire dynamics \citep{Abbottif99}
under various contributions, namely, 
\begin{equation}
\tau\frac{dV(t)}{dt}=-V(t)+V_{\text{in}}(t)+V_{\text{ext}}\left(t\right)+V_{\text{noise}}(t)+V_{0},\label{eq:general}
\end{equation}
where $V_{0}=RI_{0}$ is a constant voltage term induced by a constant
current $I_{0}$ (so that $R$ characterizes the neuron membrane resistance),
and $V_{\text{ext}}$ stands for an external well-defined signal that
we in practice implement as a sinus (in order to trace it easily).
These compete with a noise $V_{\text{noise}},$\textbf{\ }which corresponds
to uncorrelated depolarizing signals from other areas of the brain,
and we assume here that such excitatory inputs occur at times that
are Poisson distributed with mean $\mu.$ This, which also characterizes
the noise distribution broadness, will be used as a principal parameter
in our study. Furthermore, a main contribution in (\ref{eq:general})
is the total signal $V_{\text{in}}$ arriving to the given neuron
from its presynaptic (neighbor) neurons. In order to take phenomenological
account of the observed dynamic behavior of synaptic connections \citep{Torres2015},
we assume this may be written as (cf. thin lines in Fig.\ref{fig1}B)
\begin{equation}
V_{\text{in}}(t)=\left\{ \begin{array}{ll}
\!\!\left[\Theta\left(t-t_{\text{in}}\right)-\Theta\left(t-t_{\text{in}}-t_{\text{max}}\right)\right]\epsilon\tau\! & \!\text{depolarizing inputs}\\
\eta\tau\Theta\left(t-t_{\text{in}}\right)\exp\left[-\left(t-t_{\text{in}}\right)/\tau\right] & \text{hyperpolarizing inputs.}
\end{array}\right.\label{eq:2bis}
\end{equation}
Here, $t_{\text{in}}$ is the time at which the presynaptic input
occurs, the first line is for the excitatory input of amplitude $\epsilon\tau$
and duration $t_{\max}$ arriving to the neuron, and the second line
stands for the exponentially-decaying inhibitory input (decreasing
$\eta$ per unit time). $\Theta(X)$ is the Heaviside step function.

For $V_{0}=V_{\text{ext}}=V_{\text{noise}}=0,$ one may prove by exact
integration of (\ref{eq:general}) with (\ref{eq:2bis}) that the
induced depolarizing and hyperpolarizing waves generated by a single
input from an excitatory neuron and an inhibitory presynaptic one
are, respectively, 
\begin{equation}
V^{E}(t)=\left\{ \begin{array}{ll}
0\, & \text{ }t\leq t_{\text{in}}\\
\epsilon\tau_{1}\left\{ 1-\exp\left[-(t-t_{\text{in}})/\tau_{\text{1}}\right]\right\}  & \text{ }t_{\text{in}}<t\leq t_{\text{in}}+t_{\text{max}}\\
\lambda\exp\left[-(t-t_{\text{in}}-t_{\text{max}})/\tau_{\text{1}}\right] & \text{ }t>t_{\text{in}}+t_{\text{max}}
\end{array}\right.\label{eq:3E}
\end{equation}
and 
\begin{equation}
V^{I}(t)=\eta(t-t_{\text{in}})\,\exp\left[-(t-t_{\text{in}})/\tau_{\text{2}}\right],\quad\text{for }t>t_{\text{in}},\label{eq:3I}
\end{equation}
where $\lambda\equiv V^{E}\left(t_{\text{in}}+t_{\text{max}}\right)=\epsilon\tau_{\text{1}}\left[1-\exp\left(-t_{\text{max}}/\tau_{\text{1}}\right)\right]$
and $\tau_{\text{1}}\left(\tau_{\text{2}}\right)$ is $\tau$ in (\ref{eq:2bis})
for excitatory (inhibitory) inputs. That is, the absolute values of
$V^{E}(t)$ and $V^{I}(t)$ decay exponentially towards a membrane
rest value after a time $t=t_{e}$ --- being $t_{e}=t_{\text{in}}+t_{\text{max}}$
for $V^{E}(t)$ and $t_{e}\approx t_{\text{in}}+0.06$ for $V^{I}(t)$
--- with respective characteristic times $\tau_{\text{1}}$ and $\tau_{\text{2}}$.
These functions are illustrated in panel B of Fig.\ref{fig1}.

In order to reproduce the antecedent in \citep{LopesdaSilva1974}
from this formalization, one needs to discretize the above continuous
dynamics by defining instants $t_{i}=i\Delta t,\,i=1,\ldots,n,$ with
$\Delta t$ a time interval, which we assume to be $\Delta t=40\,\mu sec$
in practice. We then obtain from (\ref{eq:general}), for $V_{0}=V_{\text{ext}}=V_{\text{noise}}=0$
and denoting $V_{i}=V(t_{i}),$ that this discretely evolves under
the action of a the depolarizing and hyperpolarizing inputs, respectively,
as 
\begin{equation}
V_{i+1}=\left\{ \begin{array}{l}
\ensuremath{a_{E}V_{i}+\epsilon\Delta t\,\left[\Theta\left(i-i_{\text{in}}\right)-\Theta\left(i-i_{\text{in}}-i_{\text{max}}\right)\right]}\\
\\
\ensuremath{a_{I}V_{i}+\eta\Delta t\,\Theta\left(i-i_{\text{in}}\right)\exp\left[-\left(i-i_{\text{in}}\right)\left(1-a_{I}\right)\right]}
\end{array}\right.\label{eq:4}
\end{equation}
where $a_{E}=1-\Delta t/\tau_{\text{1}},$ $a_{I}=1-\Delta t/\tau_{\text{2}}$
and $i_{\text{in}}$ is the time step at which the presynaptic hyperpolarizing
pulse occurs, that is, $t_{\text{in}}=i_{\text{in}}\Delta t$ and
$i_{\text{max}}=t_{\text{max}}/\Delta t=100$. One may generalize
this expression to the cases of a train of $m$ depolarizing or hyperpolarizing
pulses at temporal points $i_{1},\ldots,i_{m}$ by writing, respectively:
\begin{equation}
V_{i+1}=a_{E}V_{i}+\epsilon\Delta t\,\sum_{_{k=1}}^{m}\left[\Theta\left(i-i_{k}\right)-\Theta\left(i-i_{k}-i_{\text{max}}\right)\right],\label{eq:ex}
\end{equation}
\begin{equation}
V_{i+1}=a_{I}V_{i}+\eta\Delta t\sum_{k=1}^{m}\Theta\left(i-i_{k}\right)\exp\left[-\left(i-i_{k}\right)\left(1-a_{I}\right)\right].\label{eq:in}
\end{equation}
It should be noted here that several, either depolarizing or hyperpolarizing,
waves can occur at the same time step. Also noticeable is that the
first terms in these two equations correspond to the final exponential
decreases in absolute value toward the resting value of $V$ after
the last depolarizing or hyperpolarizing pulses with characteristic
time constants $a_{E}$ and $a_{I}$, respectively. Following \citep{LopesdaSilva1974}
to prevent that the sum of depolarizing pulses in the second term
of (\ref{eq:ex}) makes the voltage $V_{i}$ to overpass its maximum
value $V_{\text{sat}},$ we introduced a factor $\left(V_{\text{sat}}-V_{i}\right)/V_{\text{sat}}$
multiplying this term. Likewise, to prevent that the sum of hyperpolarizing
pulses in the second term of (\ref{eq:in}) makes $V_{i}$ to go below
its minimum $V_{\text{min}},$we introduced a factor $\left(V_{\text{min}}-V_{i}\right)/V_{\text{min}}$.
Therefore, the final dynamics for a given neuron that receives $m$
depolarizing pulses and $l$ hyperpolarizing pulses from the presynaptic
neurons becomes 
\begin{align}
V_{i+1} & =aV_{i}+\frac{V_{\text{sat}}-V_{i}}{V_{\text{sat}}}\,\sum_{_{k=1}}^{m}\epsilon\Delta t\,\left[\Theta\left(i-i_{k}\right)-\Theta\left(i-i_{k}-i_{\text{max}}\right)\right]\nonumber \\
 & +\frac{V_{\text{min}}-V_{i}}{V_{\text{min}}}\,\sum_{k=1}^{l}\eta\Delta t\,\Theta\left(i-i_{k}\right)\allowbreak\exp\left[-\left(i-i_{k}\right)\left(1-a_{I}\right)\right]\label{eq:5}
\end{align}
where $a=a_{E}$ or $a_{I}$ depending on whether the potential $V_{i}$
after the last received pulse is either above or below $V_{\text{rest}}$.

This time evolution is conditioned by the fact that the neurons membrane
potential in the resting state is set $V_{\text{rest}}=0\,mV,$ and
not allowed either to decrease in the course of hyperpolarization
below $V_{\text{min}}=-20\,mV$ nor exceed the saturation level $V_{\text{sat}}=+90\,mV,$
both limits within the known physiological range. In fact, (\ref{eq:general})
involves the usual re-scaling $V(t)\rightarrow V(t)+60\,mV$ of the
membrane potential in actual neurons \citep{kochbook}. Concerning
the model dynamics (\ref{eq:5}), note also that, for \textit{E} neurons,
the first sum of its right-hand side is such that the times $t_{k}=i_{k}\Delta t\,$($k=1,...,m^{\prime}$)
at which the depolarizing (excitatory) inputs arrive to these\ neurons
from outside the network are Poisson distributed; such term corresponds
to $V_{\text{noise}}$ (see below). Likewise, the second sum in (\ref{eq:5})
corresponds in this case to inputs from\ \textit{I}\ neurons that
fire at times $t_{k}=i_{k}\Delta t$ ($k=1,\ldots,l$) since \textit{E}\ neurons
only receive inputs from inhibitory neurons in our network. On the
other hand, in the case of \textit{I} neurons, the first sum in (\ref{eq:5})
corresponds to contributions from \textit{E}s in the network that
fire at times $t_{k}=i_{k}\Delta t$ ($k=1,\ldots,m$), and the second
term is not occur in this case since \textit{I}\ neurons only receive
inputs from \textit{E}s in the network and are isolated from the outside.

\subsection*{Inputs}

The inputs $V_{\text{ext}}$, $V_{\text{noise}}$ and $V_{0}$ arrive
only to the \textit{E} cells since the \textit{I}s play the role in
the model of communication bridges among \textit{E}s. In particular,
we consider only non-relay interneurons, i.e., local or short-axon
ones that connect with other neurons but never with distant parts
of the brain \citep{kandelbook}. Therefore, the \textit{I}s are isolated
from external influences.

Also, trying to reflect better reality, it is assumed that $V_{\text{noise}}$
is a randomly distributed series of EPSPs corresponding to depolarization
waves, and \textit{E} cells receive only uncorrelated inputs. Our
choice for this noise is based on reports showing that often these
series of action potentials are Poisson distributed \citep{Levick1964,Fuster1965}.
The noise level parameter $\mu$ represents the mean value of action
potentials per one hundred time steps and per cell, i.e., each excitatory
cell has a probability $p_{\text{noise}}=\mu/100$ per unit time of
receiving a depolarization wave from outside. Then, to simulate a
Poisson distribution of inputs with mean $\mu$, we assume that each
\textit{E} receives random inputs from $n$ external neurons with
probability $p_{\text{noise}}/n$ of firing per time step with $n$
$(=100)$ large enough so that such binomial distribution becomes
a Poissonian one.

On the other hand, the stimulus $V_{\text{ext}}$ does not in general
refer to a sensory stimulus, given that our system can be interpreted
as a small brain module with just a few hundred neurons, and $V_{\text{ext}}$
may have electrochemical contributions from neurons outside that module.

\subsection*{Firing threshold}

The main physiological properties of \textit{E} and \textit{I} neurons
are here assumed to be the same. In particular, following known facts
\citep{kochbook}, the firing threshold of both are set at $V_{\text{th}}$
($=6\,mV$ in practice) above the resting membrane potential and,
after firing, the threshold is changed to $V_{\text{sat}}$ in order
to simulate the absolute refractory period during one hundred time
units ($t_{a}=4\,msec$). Also, to simulate the relative refractory
period once the absolute refractory period lasts, we consider that
the threshold value decreases exponentially. That is, after firing
an action potential at $t_{f}$ we have 
\[
V_{\text{th}}(t)=\left\{ \begin{array}{ll}
V_{\text{sat}} & \text{ }t_{f}<t<t_{f}+t_{a}\\
6+\left(V_{\text{sat}}-6\right)\exp\left[-\kappa\left(t-t_{f}-t_{a}\right)\right] & \text{ }t_{f}+t_{a}<t.
\end{array}\right.
\]
Here, a good fit to the typical threshold stimulus strength required
to elicit an action potential during the relative refractory period
is achieved, for example, with $\kappa=2\,msec^{-1}$. This assumption
in our model differs from the standard integrate-and-fire models \citep{Abbottif99}
which assume a constant $V_{\text{th}}$, reset the membrane voltage
at $V_{\text{rest}}$ during the absolute refractory period, and assume
lack of a relative refractory period. 
\begin{figure}[tbh]
\centering{}\includegraphics[width=12cm]{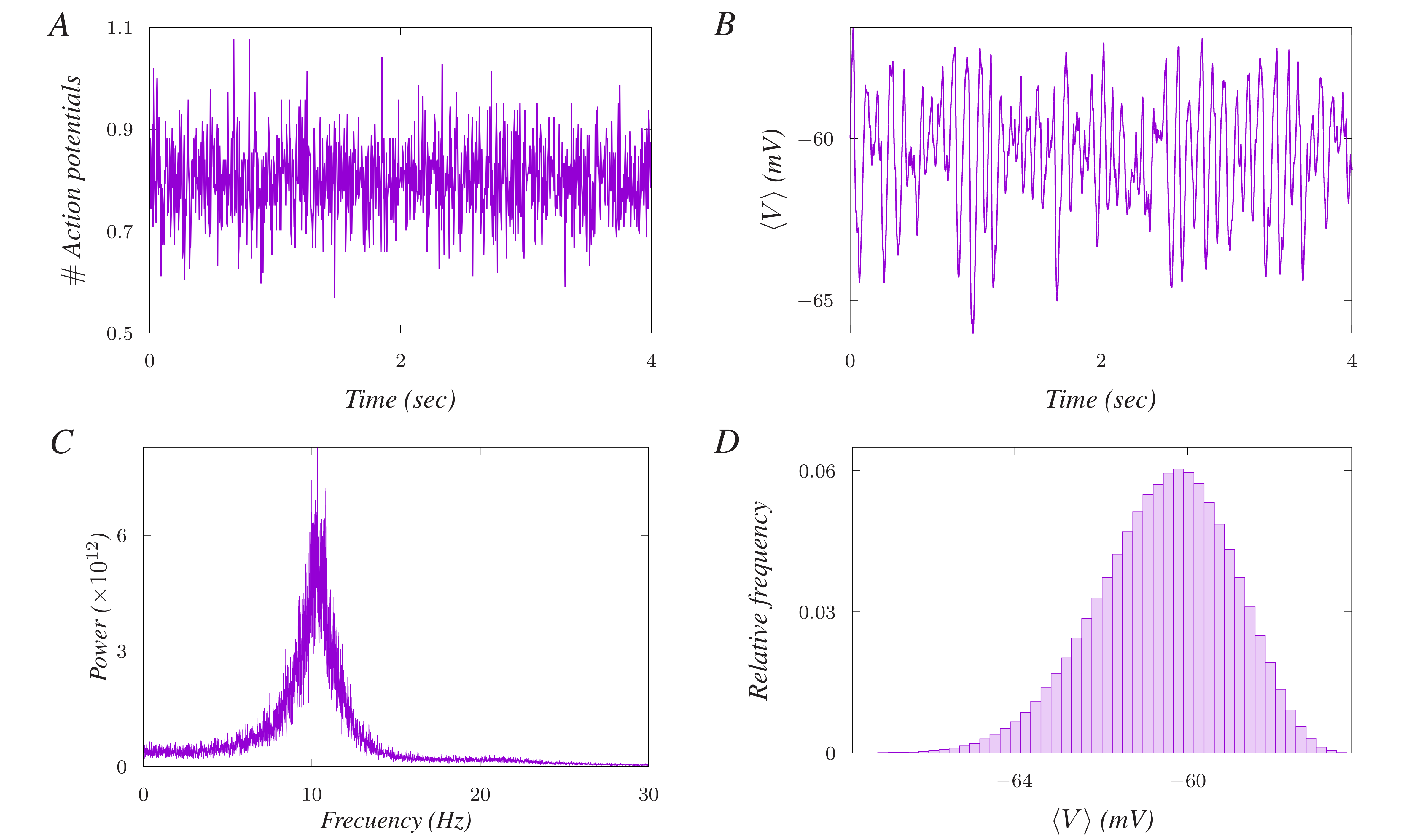}\caption{(A) Example of the noisy time series that each \textit{E}\ neuron
receives on the average from $n=100$ external \textit{E}\ neurons
from outside the network. This has a Poisson distribution of mean
$\mu=0.8$. In practice, we compute the number of external action
potentials each \textit{E}\ neuron receives each time step $t_{i}=i\Delta t$
from such distribution, and add this number to the number of depolarization
waves in the sum appearing in \eqref{eq:5}. (B) Emergent output as
measured by the average membrane potential of all the \textit{E} neurons.
Its statistical features are shown in panels (C), depicting the sharp
power spectral density, and (D), the corresponding probability distribution.}
\label{fig2}
\end{figure}

% Results and Discussion can be combined.

\section*{Results}

We monitored several dynamic variables during the network evolution
with time, including: (\textit{a}) the sum of membrane potentials
of \textit{E}\ neurons; (b)\ the same for \textit{I}\ neurons;
and (\textit{c}) the action potentials density leaving the network
via axons, i.e., the mean firing rate associated to the \textit{E}s,
i.e, $\nu(t)=(1/N_{E})\sum_{i=1}^{N_{E}}s_{i}^{E}(t)$ where $s_{i}^{E}(t)=1,0$
if the $E$ neuron is firing or not at time $t.$ Since the number
of \textit{E}s is dominant, we identify (a) with the EEG signal which,
therefore, is assumed to be the extracellular replica of the membrane
time variation. This is a sensible assumption since EEG experiments
are expected to record at a site on the scalp the summed electrical
field potentials from all cortical neurons in a certain volume of
tissue under the electrode. The fact is that a control of these quantities
shows that the model steady state is quickly attained ---typically
in around 100$\Delta t$ steps during our studies--- from any initial
condition.

Consider first the case in which $V_{\text{ext}}=V_{0}=0$ so that
the only input in Eq.(\ref{eq:general}) besides $V_{\text{in}}$
is $V_{\text{noise}}.$ When this is implemented as a Poisson distribution,
our system responds, as illustrated in Fig. \ref{fig2}, with a well-defined
rhythm wave, in spite of the wide range of frequencies in the input,
in agreement with experiments. That is, for a sufficiently large input
mean ($\mu=0.8$ in the example of Fig. \ref{fig2}) the two populations
(\textit{E} and \textit{I}) of neurons show coupled oscillations producing
collective coherent resonance, and the familiar $\alpha$-rhythm emerges.
This is revealed, for instance, by the power spectral density of the
time series for the average membrane potential over all \textit{E}
neurons, which shows a well-defined peak around $10.5$ Hz in Fig.
\ref{fig2}.

There are indications that the same simple model may generate other
types of rhythms as one varies the parameter $\mu.$ Would this be
the case, it would generalize our last observation, already reported
in \citep{LopesdaSilva1974}, along an important path as it would
indicate that all the familiar \textit{brain-rhythms} may be considered
as noise filtered by the networked system. As a matter of fact, decreasing
$\mu$ we observe that the coupling between the two populations of
neurons which induces the coherent rhythm tends to get worse. For
instance, the time series of the mean membrane potential for $\mu=0.6$
do not have the well defined periodicity nor, therefore, the acute
peak in the power spectral density in Fig. \ref{fig2}C. We shall
demonstrate below that such lack of periodicity for $\mu\sim0.6$
corresponds to a phase transition between an asynchronous condition
and a synchronous one. This fact does not show up in \citep{LopesdaSilva1974}
where a (one hundred times) larger time discretization artificially
increases synchronicity ---the same also obscures other important
facts concerning larger values of $\mu,$ as we shall illustrate below.
\begin{figure}[tbh]
\centering{}\includegraphics[width=12cm]{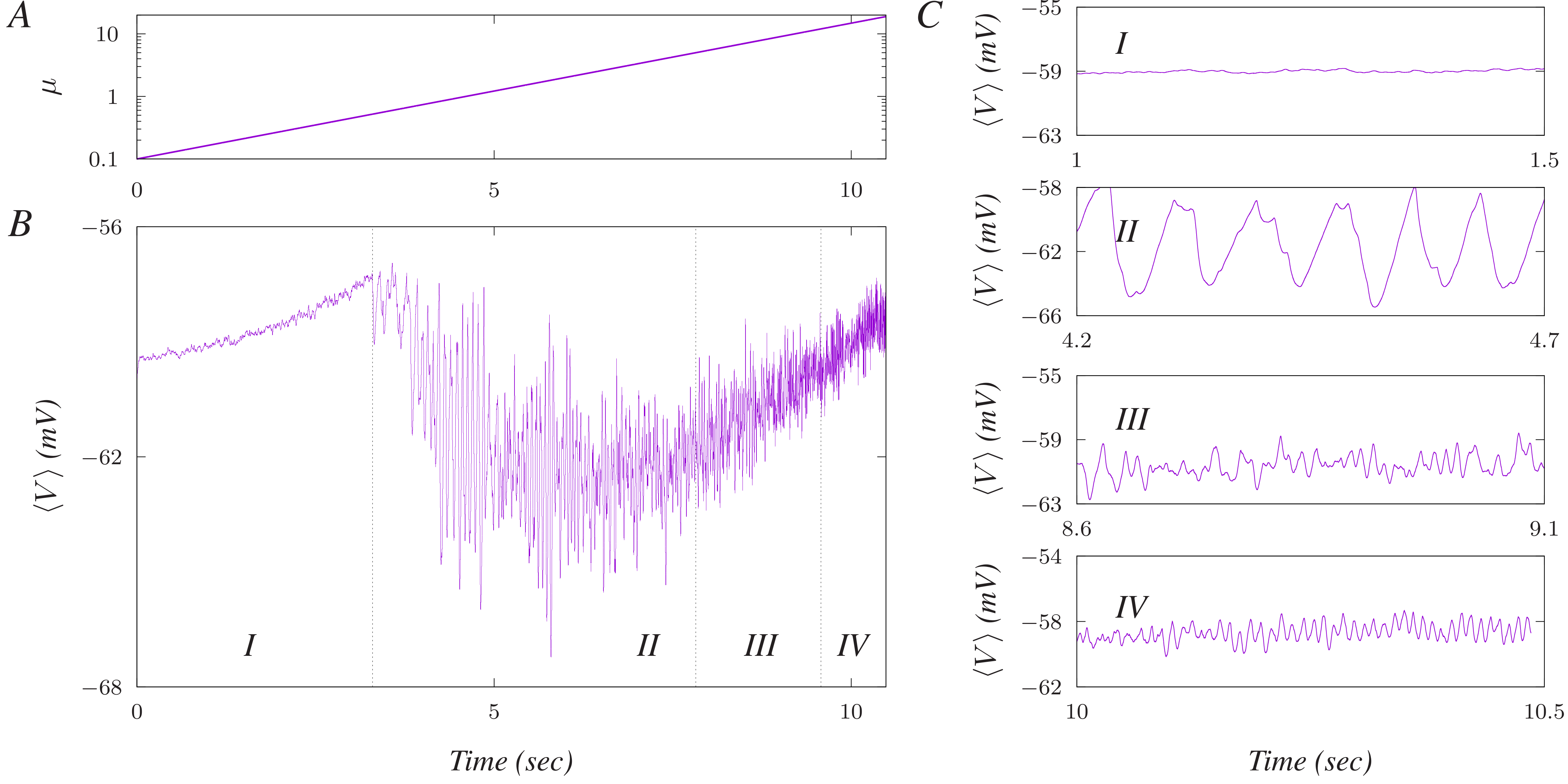}\caption{Some characteristics of the different dynamical phases that emerge
as $\mu$ is varied. Panel A shows the (linear) adiabatic temporal
variation of $\mu$ during the experiment, from $\mu=0.1$ increasing
by a factor $1.00002$ every time unit $\Delta t$. The resulting
dynamic behaviour is illustrated in panel B showing $V\left(t\right),$
and this is detailed in panel C (right) for constant $\mu$ (=0.2,
0.9, 10 and 17, respectively) within the four regions of different
behavior. Note how oscillations are too weak for $\mu\lesssim0.5$
(phase I) to speak about actual coherence resonance, while they are
clear for $\mu\gtrsim0.6,$ and coupling is observed best arround
$\mu=1.5$. Thereafter, coherence begins to decrease and the the synchronization
between the \textit{E} and \textit{I} populations decreases. Between
$\mu\approx5$ and $\mu\approx15$ there is an asynchronous phase
(III) in which frecuency cannot be defined. However, coherence and
synchrony are restored and the frecuency is well-defined again for
$\mu\gtrsim15$ (phase IV).}
\label{fig3}
\end{figure}

The new circumstance uncovered here suggested us using $\mu$ as a
control parameter, and thus explore further the emergence of brain
rhythms, which then happen to surface as characteristics of \textit{dynamic
phases}. Fig.\ref{fig3} partially illustrates the varied collective
behavior that shows up as $\mu$ is increased adiabatically in time.
This reveals that, following a rather disordered phase (I) for $\mu\lesssim0.5,$
oscillations become well defined (phase II) after $\mu\approx0.6.$
As $\mu$ is increased further, coherence is observed to decrease,
as well as the synchrony among \textit{E} and \textit{I} populations
---in particular, we observe that \textit{E}\ neurons are first
triggered simultaneously, which induces firing of \textit{I}s at the
same frequency but after a certain time lag. Asynchrony then sets
in from $\mu\approx5$ to $\mu\approx15$ (phase III), with no collective
well-defined frequency. However, coherence and synchrony with a clear
frequency are restored for $\mu\gtrsim15$ (phase IV) until $\mu>25,$
when the noise is so high that it looses any biological meaning.

To confirm these ---non-equilibrium but thermodynamic-like \citep{MD2005}---
phases, instead of slowly varying $\mu$ we also maintained the noise
constant during each simulation. Repeating this operation for different
noise values of $\mu,$ we obtained the graphs in Fig.\ref{fig3}
which happen to illustrate different types of behavior. In summary,
we may define:
\begin{description}
\item [{Phase I,}] $\mu\lesssim0.5:$ asynchrony with low activity. The
two subpopulations of neurons act almost uncoupled. No well-defined
oscillation frequency.
\item [{Phase II,}] $0.6\lesssim\mu\lesssim5:$ synchrony with broad collective
oscillations of the two subpopulations, which then oscillate coupled
at a well-defined frequency.
\item [{Phase III,}] $5\lesssim\mu\lesssim15:$ high activity with lost
of the overall coherence. Ups and downs in the average membrane potentials
of the two subpopulations are such that the excitation does not ``wait''\ for
the end of the inhibition in every period and vice-versa, so that
the periodicity and rhythm that characterize phase II is now lost.
\item [{Phase IV,}] $\mu\gtrsim15:$ synchrony, namely, the \textit{E}s
are triggered almost simultaneously, and the same with the \textit{I}s.
This is because the threshold is exceeded again in a short time (after
each firing event and its subsequent refractory period) which facilitates
synchronicity (and reduces the possibility of other type of behavior).
This synchrony goes with oscillations of the average membrane potential
with an amplitude lower than in phase II but is more regular than
these and shows a well defined oscillation frequency, as revealed
by the power spectrum. 
\begin{figure}[tbh]
\centering{}\includegraphics[width=12cm]{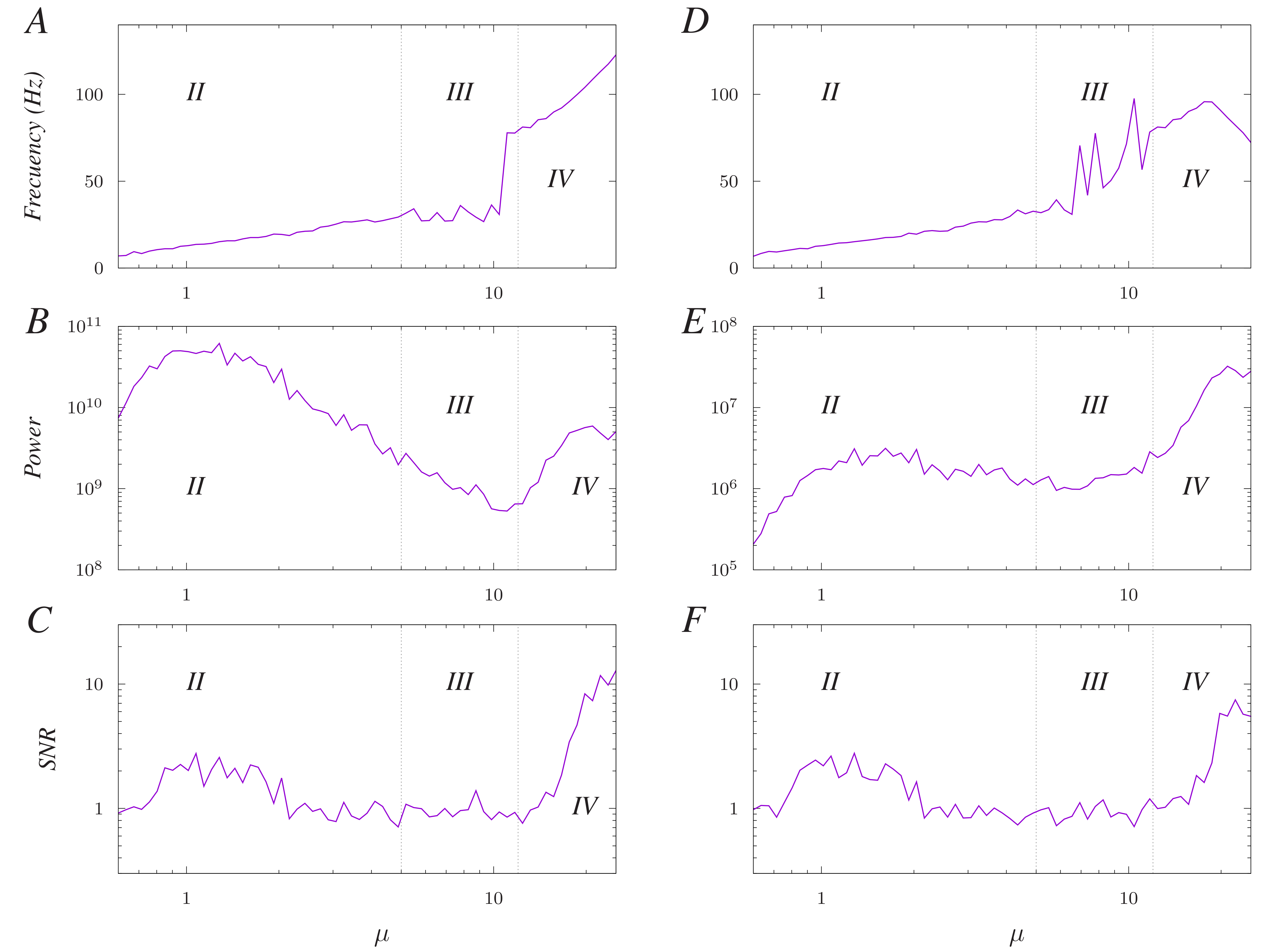}\caption{\underline{Panel A}: Frequency at the power spectra peak of time
series for the mean membrane potential as a function of $\mu$ for
$I_{0}=0$. There is asynchrony for $5\lesssim\mu\lesssim15$ (phase
III) and regions of coherent resonance before $\mu\approx5$ (phase
II) and after $\mu\approx15$ (phase IV). \underline{Panel B}: The
height of the peak in A, which is highest for phases (II and IV) due
to coherent resonance. \underline{Panel C}: Signal-to-noise ratio
(SNR). The highest values occur again for phases II and IV, and the
minimum ones during the asynchronous phases (I and III). \underline{Panels D,
E and F}: Same as in panels A, B and C, respectively, but for the time series
of the mean firing rate of the \textit{E} neurons, which confirm the
results on the left.}
\label{fig5} 
\end{figure}
\end{description}
It ensues that the familiar brain rhythms, namely, $\alpha,$ $\beta,$
$\gamma$ and ultrafast oscillations in EEG recordings from actual
awake brains, have a well-defined correspondence with these rhythmic
oscillations of the average membrane potential in the model. To clearly
uncover this, we performed extra runs lasting $2^{18}$ time steps
(equivalent to $10.5$\,s) for each of the 66 $\mu$ values in a
geometric progression starting at $\mu=0.5.$ From such time series,
we collected both the\textbf{\ }average\textbf{\ }membrane potential
and the mean firing rate of the \textit{E} population, then computed
the power spectra, and searched for a maximum on each of them. Our
main results are summarized in Fig.\ref{fig5} where panel A depicts
the frequency at which this peak occurs as a function of $\mu.$ There
is no evidence of any well defined frequency for $\mu\lesssim0.5$
(phase I, not shown), nor for $5\lesssim\mu\lesssim15$ (phase III)
which shows abrupt jumps. However, during the intermediate region
(phase II), the frequency increases from $6\,Hz$ to $25\,Hz$ ---thus
describing the spectrum of $\alpha,$ $\beta$ and $\gamma$ waves---
and, finally (phase IV), this goes from $80\,Hz$ to $130\,Hz$ ---corresponding
to high $\gamma$ and ultrafast oscillations. The same is confirmed
by time series for the mean firing of \textit{E} neurons in Fig.\ref{fig5}D.
This picture becomes even more coherent and interesting when one realizes,
as it turns out to be the case, and we develop it below, that the
passage from one behavior to a contiguous qualitatively-different
one is throughout a non-equilibrium phase transition. The system in
this way exhibits varied behavior with quite efficient features and
great economy \citep{MC2017}.

On the other hand, the peaks in Fig.\ref{fig5}B are higher in the
presence of coherent resonance, i.e., phases II and IV, than during
the asynchronous phases I (not shown since not a clear peak develops
in fact here) and III. The behavior is similar for the mean firing
rate in Fig.\ref{fig5}E. It also interests the $\mu$ variation of
the signal-to-noise ratio (SNR) at the power spectra peak, i.e., its
height divided by the average in a small range around. Even more clear
than the spectra peaks, the SNR shows maxima if coherence occurs (II
and IV) and goes to minima in the asynchronous phases (I and III).
The same is confirmed by the power spectrum of the time series for
both the membrane potential in Fig.\ref{fig5}C and the mean firing
rate in Fig.\ref{fig5}F. Specifically, the maximum coherence value
is achieved in both cases around $\mu\simeq1.3$ (within phase II)
and for $\mu\gtrsim20$ (within phase IV), and it is also noticeable
that the SNR maximum, for both the global membrane potential and the
mean firing rate, is higher for phase IV than for phase II. 
\begin{figure}[tbh]
\centering{}\includegraphics[width=12cm]{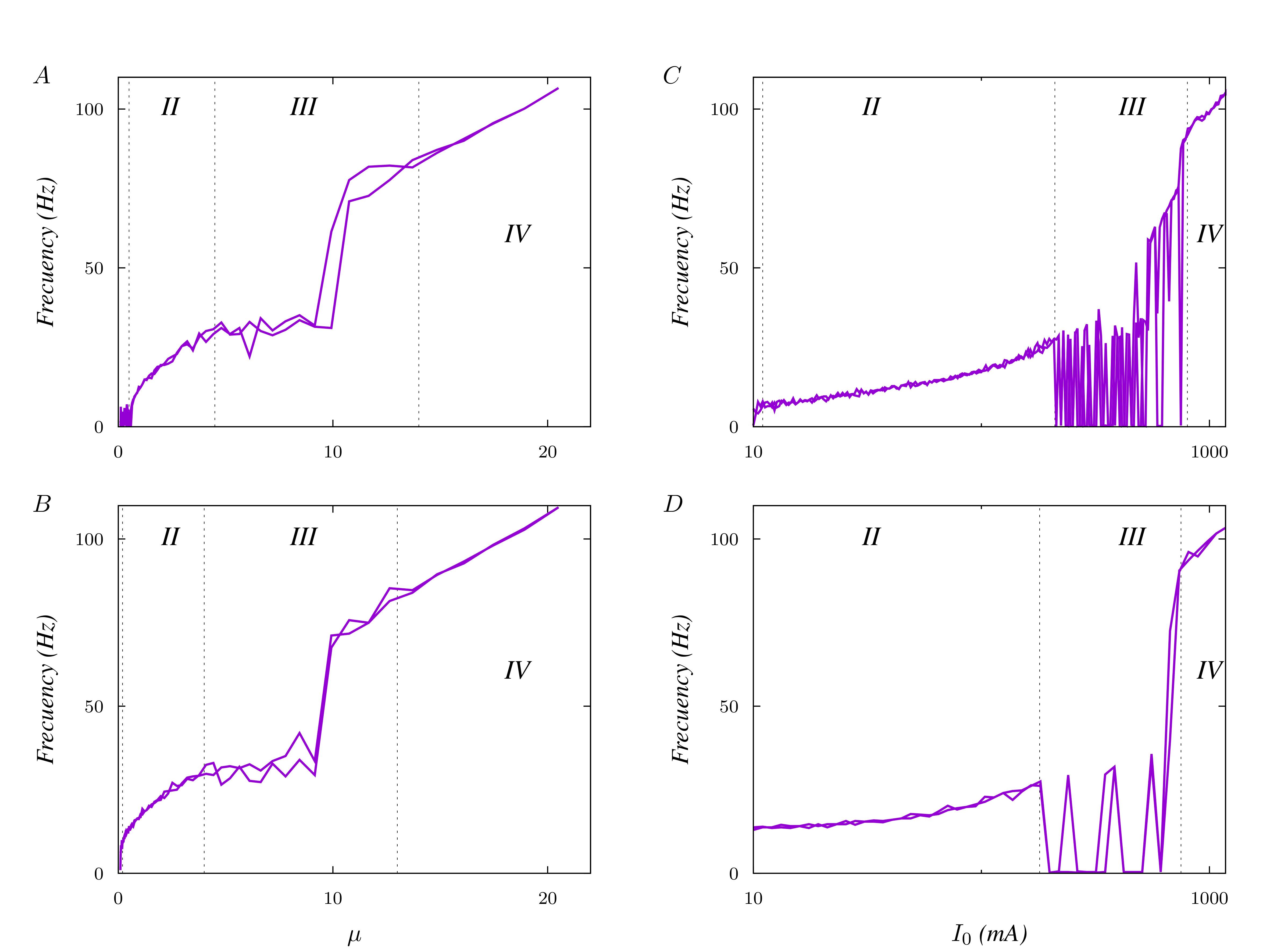}\caption{Study of hysteresis as a function of $\mu$ (left) and $I_{0}$ (right).
\underline{Panel A}: Frequency at which the peak of the power spectra
for the mean membrane potential occurs (as $\mu$ is increased and
decreased adiabatically with $I_{0}=0$). The two curves superimpose
where the frequency is well defined. \underline{Panel B}: The same
but for $I_{0}=50,$ confirming the phases in A, but shifted to the
left. \underline{Panel C}: The frequency in A but as a function of
the constant input $I_{0}$ for $\mu=0.5,$ which confirms the same
phases and shows that $210\leqslant I_{0}\leqslant750$ is an asynchronous
region in which frequency is not well-defined. \underline{Panel
D}: Same as in panel C but for $\mu=1$ showing the same but with changes
now shifted to the left relative to panel C because $\mu$ is now
higher.}
\label{fig6} 
\end{figure}

The above suggests a great interest in characterizing the transition
regions separating qualitatively different behaviors as one varies
$\mu$ and $I_{0}.$ Particularly, there is interest in the transition
between phases III and IV. Fig.\ref{fig5}A, for instance, reveals
that this is sharp, suggesting a thermodynamic-like discontinuous
phase transition. To address this, we run our system during 10s for
each $\mu$ value, as we varied adiabatically this parameter in geometric
progression while keeping $I_{0}=0.$ We retained the final state
of all the neurons in each run to serve as the initial state for the
run at the next noise value, which only differs in a small percentage
from the previous one. Once the maximum $\mu$ is reached, the process
is reverted, keeping again each final state as the initial one during
this noise reduction process. The resulting hysteresis cycle around
transitions III$\leftrightarrow$IV is shown in Fig.\ref{fig6}A,
which confirms the discontinuous first-order-like nature of the phase
transition. Such (even small) hysteresis seems to reflect that the
frequency of the global oscillations is not well-defined in phase
III; in fact, this shows no clear peak in the power spectra, and the
maximum we use to compute hysteresis can depend on the run conditions.
However, when the frequency is well-defined, the round-trip curves
superimpose. We obtain similar results for $I_{0}=50$ in Fig.\ref{fig6}B,
but with the phase changes somewhat shifted to the left.

The fact that our model shows the same qualitative behavior or phases
within a wide ample range of $I_{0}$ values suggests that its behavior
is robust to the type of input, and we confirmed this by moving $I_{0}$
adiabatically for $\mu=0.5$ (Fig.\ref{fig6}C) and $\mu=1$ (Fig.\ref{fig6}D).
Note that phase I is not shown, since for $\mu=1$ the system is at
phase II even for $I_{0}=10,$ that phase III occurs for $180\leq I_{0}\leq700$
and that, as expected, the phase changes are shifted to the left relative
to Fig.\ref{fig6}C because $\mu$ is now higher. The conclusion is
that the system is sensible to the total current arriving to the network
but not to the type of input. In other words, increasing the noise
and $I_{0}$ tends to increases the excitability of both neuron populations
but the emergent behavior is rather due to the complex interplay between
the activity of \textit{E} and \textit{I} populations.

\subsection*{Stochastic resonance as a detector of phase transitions in EEG activity}

\begin{figure}[h!]
\centering{}\includegraphics[width=12cm]{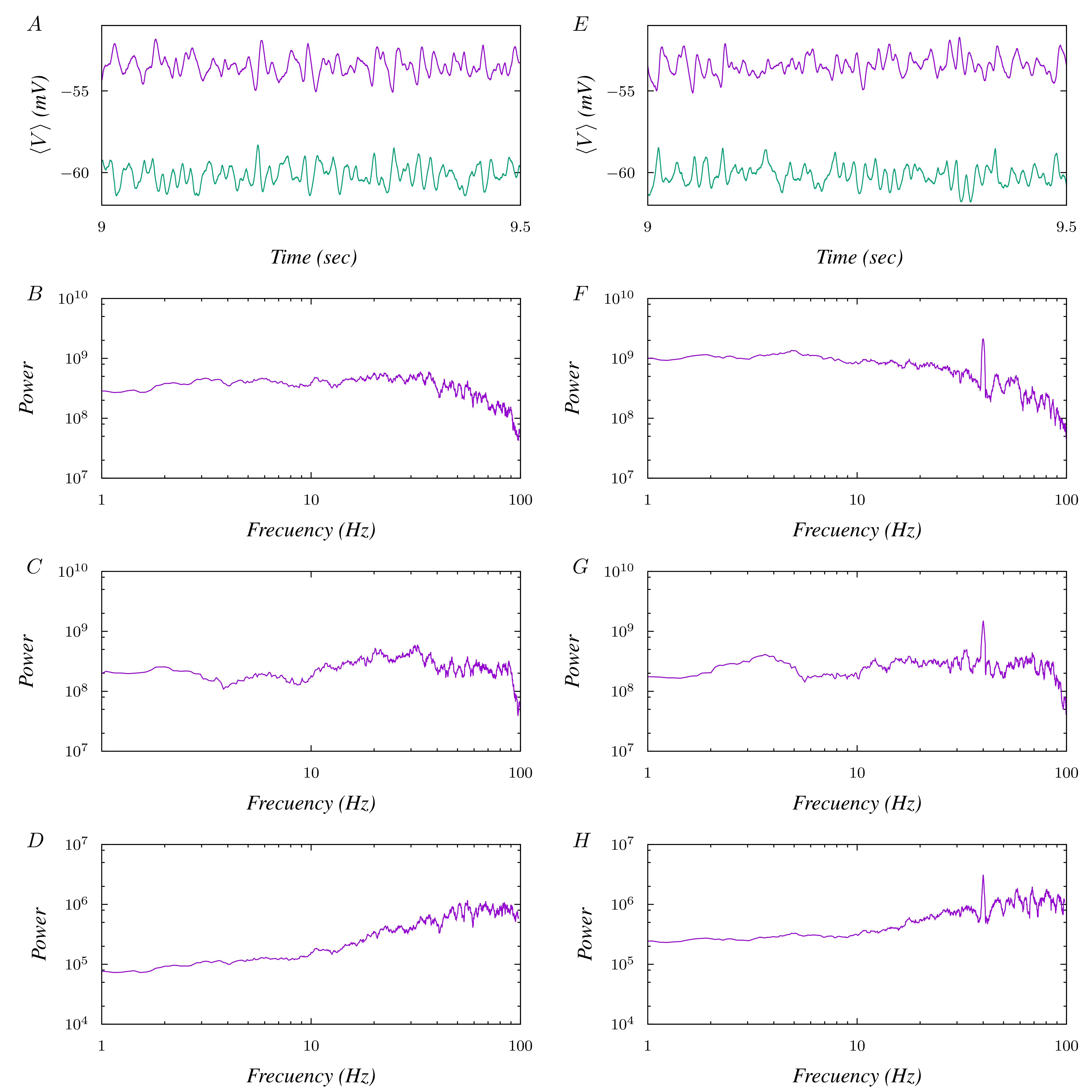}\caption{Case of $\mu=10$ showing the effect of adding (right column) a small
signal $V_{\text{ext}}=d\,\sin2\pi f\,t$ to each $E$ neuron. The
left column is for $d=0$ while the right one is for $d=25$ and $f=40\,Hz$.
Top panels show the mean membrane potential of \textit{E} (violet
line) and \textit{I} (green line) neurons with (panel E) and without
(panel A) the external signal. Panels B and F depict the corresponding
(membrane potential) power spectra for the \textit{I} neurons in both
situations, and panels C and G show the same for \textit{E} neurons.
At the bottom, panels D and H illustrate the corresponding power spectra
of the mean firing rate for \textit{E} neurons. The signal only slightly
modifies dynamics but a clear peak emerges at frequency $f$. }
\label{fig7}
\end{figure}

We also checked the case of a weak input $V_{\text{ext}}=d\,\sin(2\pi ft)$
with small $d$ to the neural network, instead of $V_{\text{ext}}=0$
as above. In general, even relatively small values of $d$ induce
a new maximum at frequency $f$ in the power spectra, as shown in
Fig.\ref{fig7} (right column).

The emergent peak here ---which happens to stand out more or less
depending on the values for $d$ and $\mu$--- reveals the existence
of the so-called \textit{stochastic resonance} (SR) phenomenon \citep{SRreview98}.
That is, the propagation of a weak signal is enhanced at certain intermediate
level of noise while it is generally obscured at lower and higher
levels of noise. The SNR in the power spectra consequently increases
at those moderate values of the noise. As it was already shown \citep{Torres2015},
this is just a consequence of the great susceptibility the cooperative
system exhibits in a region in which a phase transition occurs, so
that it provides a simple method to detect changes of qualitative
behavior in these types of systems.

\begin{figure}[h!]
\begin{centering}
\includegraphics[width=12cm]{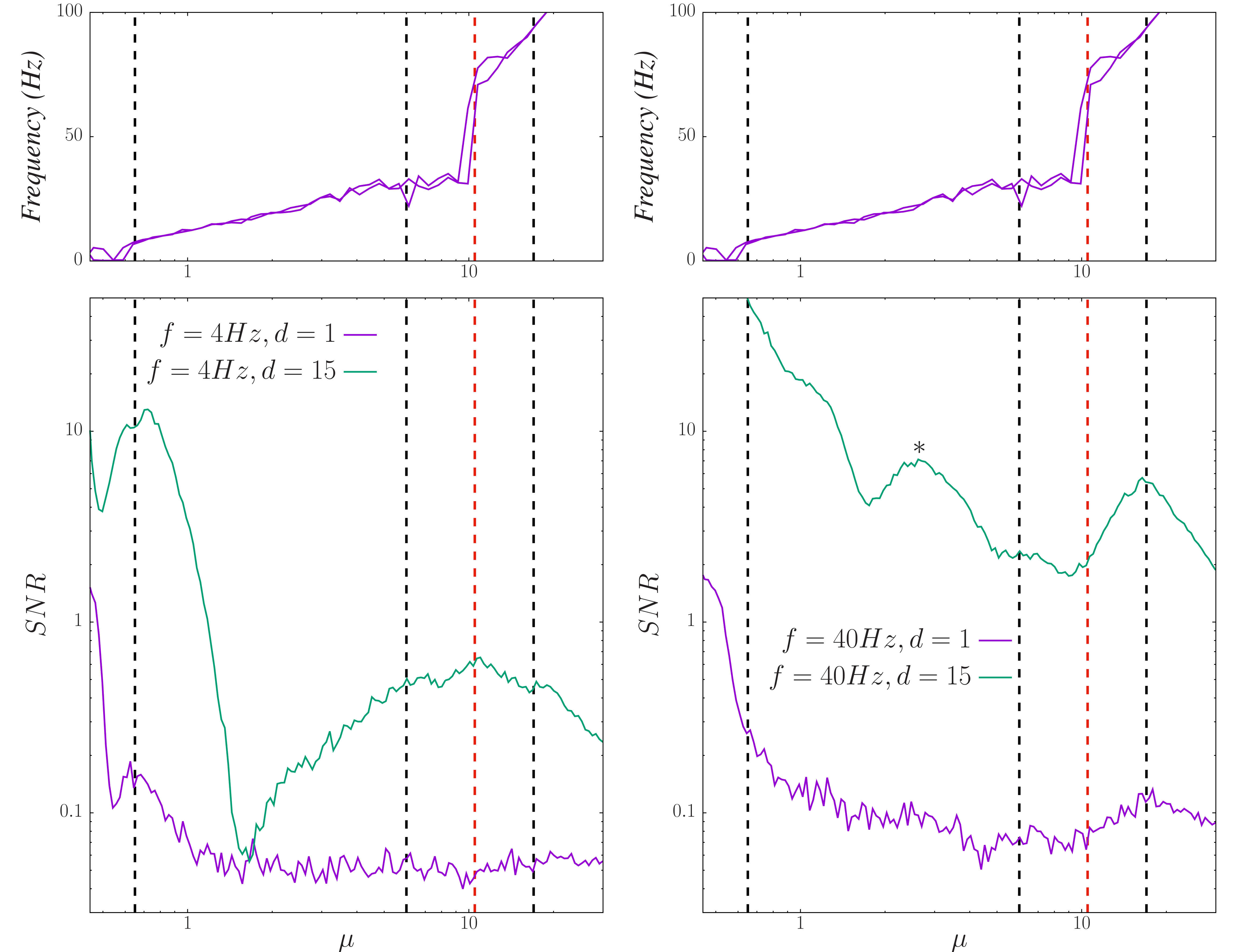}
\par\end{centering}
\caption{Emergence of \textit{stochastic resonance} in the system in figure
\ref{fig1}, for $I_{0}=0,$ when a weak sinusoidal input $V_{\text{ext}}$
of low frequency ($f\sim4Hz$) (left graph) and high frequency ($f\sim40Hz$)
(right graph) affects each $E$ neuron. SR peaks appear around the
phase transition points (vertical dashed lines at $\mu=0.6,\,6$ and
$16$) depicted in the top panels. Note in the left graph that the
jump corresponding to the change of behaviour in simulations between
phases IV and III (see top panel) appears around $\mu\approx10$ (see
red vertical dashed line) that coincides with the larger resonance
peak for large level of noise. Secondary resonance peaks occurs around
this maximun for $\mu\approx6$ and $16.$ In the right panel, however,
such maximun does not show. Also, the low noise resonance peaks around
$\mu\approx0.6$ is neither appearing and the only ones are those
around $\mu\approx6$ (poorly seeing) and $16$ (very clearly depicted).
Different \emph{SNR} curves here were obtained after averaging over
100 trials and computing the power spectra over a time series of $2^{18}ms$
for each trial.}
\label{fig8}
\end{figure}

A general evidence of SR phenomena in the system is illustrated in
Fig.\ref{fig8} for $I_{0}=0,$ showing the\emph{ signal to noise
ratio ($SNR$)} as a function of the noise level $\mu$ for both low-frequency
($\sim4Hz$) and high-frequency ($\sim40Hz$) inputs signals. In agreement
with the interpretation of stochastic resonance in \citep{Torres2015},
here we observe how SR peaks develop around the phase transitions
described above. For low-frequency signals (left graph) there are
clear maximum at $\mu\approx0.6,$ $6$ and $16$ corresponding to
the phase transitions I$\leftrightarrow$II, II$\leftrightarrow$III
and III$\leftrightarrow$IV, respectively. The $SNR$ also shows a
peak for $\mu\approx10$ which corresponds to the level of noise at
which finite-size jumps between III$\leftrightarrow$IV occurs in
simulations. The emergence of this peak can be explained assuming
that noise makes that these finite-size jumps of activity between
both phases can be driven by the weak stimulus, so an amplification
of the weak signal occurs at such noise level. Then, we expect that
such peak will disappear as the network size is increased which will
be an indication that the transition III$\leftrightarrow$IV is of
first-order type as simulations seams to indicate (see top graphs
in Fig. \ref{fig8}). For high-frequency signals ($\sim40Hz$), only
the transitions II$\leftrightarrow$III and III$\leftrightarrow$IV
are clearly marked by stochastic resonance peaks around $\mu\approx6$
and $16,$ the first hardly distinguishable and the last one very
clear. The peak around $\mu\approx0.6$ is not appearing due to the
fact that system oscillations at such level of noise at a natural
frequency of alpha range around $10Hz$ or less, which is very small
compared with the weak signal stimulation frequency ($40Hz$). This
is incompatible with the emergence of the SR where the stimulation
frequency must be very low compared with the intrinsic oscillation
frequency of the system. Note that this impediment does not occur
for the the resonance peak around $\mu=17$ since for this case the
intrinsic oscillation frequency of the system is around $75Hz$ or
larger which is bigger that the stimulation frequency of $40Hz$,
so conditions for the emergence of SR still hold. The transition II$\leftrightarrow$III
occur around $\mu\approx0.6$ with system oscillations of frequency
around the stimulation frequency, so this is the reason why the SR
peak around $\mu\approx0.6$ is not so clearly depicted. Also remarkable
in this high-frequency stimulation case is the presence of an additional
resonance peak around $\mu\approx2.5$ (marked with ``{*}'' in the
figure) which corresponds with a range of frequencies$\sim25-30$$Hz$,
and it could indicate the exact limit between $\beta$ (with intrinsic
frequency between $12$ to $30Hz$) and $\gamma$ brain waves (with
intrinsic frequency larger than $30Hz$) as experimental psychologists
and neuroscientists have widely described (see for instance \citep{beta0,beta12,beta2}).
This overall behaviour should also be discernible in actual EEG experiments.

\section*{Discussion}

We here present an extension, and formalization according to recent
familiar standards, of a model for the generation of brain $\alpha$
rhythms \citep{LopesdaSilva1974} which provides a simple and well-defined
scenario also for other types of brain waves. In addition to signals
from other neurons $\left(V_{\text{in}}\right),$ and from outside
the network ---which are globally portrayed here as a Poisson noise
$\left(V_{\text{noise}}\right)$ which is characterized by the parameter
$\mu$--- our model Eq.(\ref{eq:general}) includes a constant current
$I_{0}$ and a small external input signal $V_{\text{ext}}$. Our
main findings may be summarized as follows:
\begin{itemize}
\item Previous results \citep{LopesdaSilva1974} are confirmed and, using
realistic, smaller time steps, we describe lower degrees of coherence
and real levels of synchronicity.
\item In this way, we identify four different ``phases''\ or qualitative
types of dynamic behavior in the model. As $\mu$ is increased, this
exhibits oscillations that are too weak in amplitude so that any coherence
is precluded (phase I), coherent resonance and synchrony (phase II),
asynchrony showing abrupt jumps in the corresponding frequency curves
(phase III), and coherence and synchrony with a well-defined frequency
again (phase IV).
\item In phase II, our system precisely includes the frequency spectrum
of $\alpha,$ $\beta$ and low $\gamma$ waves of actual EEG recordings,
and phase IV covers the frequencies corresponding to high $\gamma$
and ultrafast oscillations.
\item The highest coherent resonance, as revealed by the power spectra peak
and the corresponding SNR, is for phases II and IV, while the lowest
one occurs in the asynchronous phases I and III. 
\item The average amount of electrical impulses arriving to the network
per unit of time ---that we parametrize as $\mu$--- is essential
to characterize the different phases, more than the nature, either
constant or noisy, of the input. 
\item \textit{Stochastic resonance} \citep{Torres2015,MC2017} is revealed,
e.g., by SNR, locating changes of qualitative behavior when the system
receives a signal. We confirm that this fact may provide a powerful
tool to investigate phase transitions in mammals' and other brains
using simple techniques such as EEG recordings or simple experiments
as devised in \citep{Torres2015}. 
\end{itemize}
The above picture indicates, on one hand, that a single mechanism
is behind the familiar brain rhythms and, on the other, that such
waves are related to the general phenomena of non-equilibrium phase
transitions, where a system is known to be highly susceptible, efficient
and adaptable \citep{MD2005,MC2017}. This is compatible with specific
mechanisms that might act during the generation of brain oscillations\textcolor{blue}{{}
}while cognitive functions occur. However, a one-to-one correspondence
between different type of brain oscillations and cognitive functions
cannot be stablished in fact, there are many more different cognitive
processes than types of brain waves \citep{herrmann2016}. It seems
sensible to assume that similar brain waves in the same frequency
band can contribute to different cognitive functions depending on
the particular brain area in which they originated and on their particular
temporal features \citep{herrmann2016}. For example \citep{Miller200718},
while local synchronization during visual processing evolves in the
$\gamma$ range, synchronization between neighboring temporal and
parietal cortex during multi-modal semantic processing may evolve
in a lower $\beta$ (12-18 Hz) range, and long range fronto-parietal
interactions during working memory retention and mental imagery in
the $\theta$ (4-8 Hz) and $\alpha$ (8-12 Hz) ranges. That is, a
relationship may exist between functional integration and synchronization
frequency which could be due to conduction delays in long corticortical
axons ---up to several tens of ms for conduction distances of $\sim$10
mm--- and convert $\gamma$ to $\beta$ oscillations (with cycle
times ranging from $30$ to $70$ ms). The same process for more widely-dispersed
interactions could produce activity in the active cortex in the $\alpha$
range (cycle time 77--125 ms) or even in the $\theta$ range. To
our knowledge, however, these details have yet been poorly demonstrated
and the argument requires that all axonal connections of a given network
were approximately the same length, which is a too strong assumption
for regions of arbitrary extension. In addition, electrical stimulation
of V1 induces enhanced $\gamma$-band activity in V4, whereas V4 stimulation
induces enhanced $\alpha\mbox{{-}}\beta$-band activity in V1 \citep{michalareas2016},
when it is supossed that the conduction delays are aproximately the
same from V1 to V4 than from V4 to V1. Recognizing that the presence
of conduction delays may importantly complicate the network dynamics
\citep{lee2009}, and that brain oscillations could be related to
many body oscillations \citep{klimesch2018} -- as heart rate, heart
rate variability, breathing frequencies, fluctuations in the BOLD
signal, and others -- our proposal does not require any hypothesis
concerning conduction times or too speculative assumptions concerning
the coupling of the brain activity with any type of body oscillations.\textcolor{blue}{{}
}From a different point of view, given the modular structure of the
brain \citep{Bullmore2009}, we may imagine small networks with a
great internal connectivity, each as ours here and perhaps in some
of the dynamic phases we have described subject to an input. Furthermore,
it is sensible to assume that, in a large region of interconnected
neurons, an input from other modules will not affect all the neurons,
since otherwise it might induce an anomalous high physiological level
of activity. On the average, one should expect our parameter $\mu$
to be low and only high inputs eventually reaching small local regions.
Within this scenario, our model suggests that large synchronized regions
receive small inputs, and therefore will oscillate in the $\alpha$
regime, while small local synchronized regions receiving a large input
will oscillate synchronously in the $\gamma$ range. Our scenario
is thus compatible with the one in \citep{vonStein2000}.

Also, we mention that some authors associate consciousness with coherent
$\gamma$ oscillations in different parts of the brain, and thus explain
episodes of attention \citep{Crick-Koch1990}. In the light of our
results, we can hypothesize that the transition III$\rightarrow$IV
could be related to the emergence of awareness of memories associated
with the modules that reach the corresponding input, a hyphothesis
that could be tested experimentally. In fact, we could include all
the 40-70 Hz frequencies in phase IV choosing adequately model parameter
values. In particular, our network model may easily involve a small
random delay of small variance in all the connections, a topology
different that the one in Fig.\ref{fig1}A, and/or vary the parameters
of the EPSP and IPSP waves in Fig.\ref{fig1}B to achieve this. Other
theories of consciousness, as the Integrated Information Theory \citep{oizumi2014}
and its continuous dynamical system version \citep{galadi2018} are
also consistent with our scenario in which one may have two phases
with very different levels of activity, both with a synchronicity
that facilitates the communication with other mechanisms, and our
phases II and IV would be equivalent to the ``off''\ and ``on''\ states
in this theory.

Finally, our findings here can also be useful to design appropriate
deep learning algorithms based on SR \citep{sra2016,deeplearningreview2019}
which might optimally learn feature representation in the presence
of noise in actual EEG recordings, or to identify phase transitions
in the brain from raw data in EEG time series \citep{zhang2018}

\section*{Acknowledgments}

We acknowledge the Spanish Ministry for Science and Technology and
the \textquotedblleft Agencia Española de Investigación\textquotedblright{}
(AEI) for financial support under grant FIS2017-84256-P (FEDER funds).




\begin{thebibliography}{10}
\bibitem{chialvo_eguiluz05}V.~M. Eguiluz, D.~R. Chialvo, G.~A.
Cecchi, M.~Baliki, and A.~V. Apkarian. \newblock Scale-free brain
functional networks. \newblock {\em Phys. Rev. Lett.}, 94:018102,
2005.

\bibitem{Hesse2014}J.~Hesse and T.~Gross. \newblock Self-organized
criticality as a fundamental property of neural systems. \newblock\emph{Frontiers
in Systems Neuroscience}, 8:166, 2014.

\bibitem{Torres2015}J.~J. Torres and J.~Marro. \newblock Brain
performance versus phase transitions. \newblock {\em Scientific
Reports}, 5:12216 EP --, 2015.

\bibitem{MC2017}J. Marro and D. R. Chialvo, \textit{La Mente es Crítica
--- Descubriendo la admirable complejidad del cerebro}, EUG (University
of Granada Press), Granada 2017 (an English new version of the contents
of this book is now in progress).

\bibitem{munoz2018}M. A . Muñoz, \newblock Colloquium: Criticality
and dynamical scaling in living systems, \newblock \emph{Review of
Modern Physics.} 90(3), 031001, 2018.

\bibitem{cocchi2017}L. Cocchi, L. L. Gollo, A. Zalesky, M. Breakspear.
\newblock Criticality in the brain: A synthesis of neurobiology,
models and cognition. \newblock \emph{Progress in Neurobiology, }158,
132-152, 2017.

\bibitem{gollo2017}L. L. Gollo. \newblock Coexistence of critical
sensitivity and subcritical specificity can yield optimal population
coding. \newblock \emph{Journal of The Royal Society, Interface,
}14 (134), 2017.

\bibitem{LopesdaSilva1974}F.~H. Lopes~da Silva, A.~Hoeks, H.~Smits,
and L.~H. Zetterberg. \newblock Model of brain rhythmic activity.
\newblock {\em Kybernetik}, 15(1):27--37, 1974.

\bibitem{MD2005}J. Marro and R. Dickman, \textit{Nonequilibrium Phase
Transitions in Lattice Models, }Cambridge University Press, Cambridge
2005.

\bibitem{SRreview98}L.~Gammaitoni, P.~Hänggi, P.~Jung, and F.~Marchesoni.
\newblock Stochastic resonance. \newblock {\em Review of Modern
Physics.}, 70:223--287, Jan 1998.

\bibitem{zhang2018}W. Zhang, J. Liu, T.-Ch. Wei \newblock Machine
learning of phase transitions in the percolation and XY models.\newblock
arXiv:1804.02709v1, 2018.

\bibitem{sra2016}S. Ikemoto ; F. Dalla-Libera ; K. Hosoda, \newblock
Stochastic resonance induced continuous activation functions in a
neural network consisting of threshold elements\newblock 2016 International
Joint Conference on Neural Networks (IJCNN), IEEE publisher, pp 2603--2608,
2016

\bibitem{deeplearningreview2019}Y. Roy, H. Banville, I. Albuquerque,
A. Gramfort, T. H. Falk, J. Faubert. \newblock Deep learning-based
electroencephalography analysis: a systematic review.\newblock arXiv:1901.05498v2,
2019.

\bibitem{Tombol1967}T.~Tömböl. \newblock Short neurons and their
synaptic relations in the specific thalamic nuclei. \newblock \emph{Brain
Research}, 3(4):307--26, 1967. \newblock

\bibitem{Braitenberg1998}V.~Braitenberg and A.~Schüz. \newblock
\emph{Cortex: Statistics and Geometry of Neuronal Connectivity}. \newblock
Springer, 1998.

\bibitem{Markram2004}H.~Markram, M.~Toledo-Rodriguez, Y.~Wang,
A.~Gupta, G.~Silberberg, and C.~Wu. \newblock Interneurons of
the neocortical inhibitory system. \newblock {\em Nature Review
of Neuroscience}, 5(10):793--807, Dec 2004.

\bibitem{Okun2009}M.~Okun and I.~Lampl. \newblock {B}alance
of excitation and inhibition. \newblock {\em Scholarpedia}, 4(8):7467,
2009.

\bibitem{Abbottif99}L. F. Abbott. \newblock Lapicque's introduction
of the integrate-and-fire model neuron (1907). \newblock\emph{\ Brain
Research Bulletin,} 50(5-6):303--304, 1999.

\bibitem{kandelbook}E.~R. Kandel, J.H. Schwartz, and T.~M. Jessell.
\newblock {\em Principles of Neural Science}. \newblock McGraw-Hill,
4th edition, 2000.

\bibitem{Levick1964}W.~R. Levick and W.~O. Williams. \newblock
Maintained activity of lateral geniculate neurones in darkness. \newblock
\emph{J Physiol}ogy, 170:582--597, 1964.

\bibitem{Fuster1965}J.~M. Fuster, A.~Herz, and O.~D. Creutzfeldt.
\newblock Interval analysis of cell discharge in spontaneous and
optically modulated activity in the visual system. \newblock {\em
Archives Italiennes de Biologie}, 103(1):159--177, 1965.

\bibitem{kochbook}Ch. Koch. \newblock {\em Biophysics of Computation:
Information Processing in Single Neurons}. \newblock Oxford University
Press, 1999.

\bibitem{beta0}R. E. Dustman, R. S. Boswell, P. B. Porter. Beta Brain
Waves as an Index of Alertness. \emph{Science }137(3529): 533--534,
1962.

\bibitem{beta12}G. Buzsáki. Rhythms of the Brain. New York: Oxford
University Press. p. 4. 2006.

\bibitem{beta2}J. D. Kropotov. Beta Rhythms. In: Quantitative EEG,
Event-Related Potentials and Neurotherapy, Academic Press, San Diego,
2009, Pages 59--76 2009.

\bibitem{herrmann2016} C.S. Herrmann, D. Strüber D, R.F. Helfrich,
A.K. Engel. \newblock EEG oscillations: from correlation to causality.
\newblock {\em International Journal of Psychophysiology}, 103:
12--21, 2016.

\bibitem{Miller200718}R.~Miller. \newblock Theory of the normal
waking eeg: From single neurones to waveforms in the alpha, beta and
gamma frequency ranges. \newblock {\em International Journal of
Psychophysiology}, 64(1):18 -- 23, 2007.

\bibitem{michalareas2016} G. Michalareas, J. Vezoli, S. Pelt, J.
Schoffelen, H. Kennedy, P. Fries. \newblock Alpha-beta and gamma
rhythms subserve feedback and feedforward influences among human visual
cortical areas.\newblock \emph{Neuron} 89: 384--397, 2016.

\bibitem{lee2009}W.~S. Lee, E.~Ott, and T.~M. Antonsen. \newblock
Large coupled oscillator systems with heterogeneous interaction delays.
\newblock\emph{Physical Review Letters}, 103:044101, Jul 2009.

\bibitem{klimesch2018} W. Klimesch \newblock The Frequency Architecture
of Brain and Brain-Body Oscillations: An Analysis. \newblock \emph{European
Journal of Neuroscience, }48(7): 2431--2453, 2018.

\bibitem{Bullmore2009}E.~Bullmore and O. Sporns. \newblock Complex
brain networks: graph theoretical analysis of structural and functional
systems. \newblock {\em Nature Reviews of Neuroscience}, 10(3):186
--198, 2009.

\bibitem{vonStein2000}A.~von Stein and J.~Sarnthein. \newblock
Different frequencies for different scales of cortical integration:
from local gamma to long range alpha/theta synchronization. \newblock\emph{International
journal of psychophysiology }, 38(3):301--13, 2000.

\bibitem{Crick-Koch1990}F, Crick and Ch. Koch. \newblock Towards
a neurobiological theory of consciousness. \newblock {\em Seminars
in the neurosciences}, 2:263 -- 275, 1990.

\bibitem{oizumi2014}M.~Oizumi, L.~Albantakis, and G.~Tononi. \newblock
From the phenomenology to the mechanisms of consciousness: Integrated
information theory 3.0. \newblock {\em PLoS Computational Biology},
10(5):1--25, 05, 2014.

\bibitem{galadi2018} F. J. Esteban, J. A. Galadí, J. A. Langa, J.
R. Portillo, and F. Soler-Toscano. \newblock Informational structures:
A dynamical system approach for integrated information. \newblock
\emph{PLoS Computational Biology,} 14(9): , 2018.
\end{thebibliography}


%\begin{thebibliography}{00}


%\bibitem{}

%\end{thebibliography}

\end{document}